\documentclass[12pt]{iopart}

\usepackage{amssymb}
\usepackage{mathrsfs}
\usepackage[latin1]{inputenc}

\newtheorem{theorem}{Theorem}

\newcommand{\U}{\mathcal{U}}
\newcommand{\N}{\mathcal{N}}
\newcommand{\p}{\mathcal{P}}
\newcommand{\n}{{\bf{n}}}
\newcommand{\mm}{{\bf{m}}}
\newcommand{\K}{{\bf{K}}}

\newcommand{\M}{\mathcal{M}}
\newcommand{\R}{\mathbb{R}}
\newcommand{\G}{\mathscr{G}}
\newcommand{\lmd}{\lambda}
\newcommand{\sd}{Schr\"{o}dinger }
\newcommand{\dd}{{\rm d}}
\newcommand{\ep}{{\varepsilon}}

\begin{document}

\title[Characterization of the Critical Submanifolds]{Characterization of the Critical Submanifolds in Quantum Ensemble Control Landscapes}

\author{Rebing Wu and Herschel Rabitz and Michael Hsieh}

\address{Department of Chemistry, Princeton University, Princeton, New Jersey 08544, USA.}
\ead{rewu@princeton.edu}
\begin{abstract}
The quantum control landscape is defined as the functional that
maps the control variables to the expectation value of an
observable over the ensemble of quantum systems. Analyzing the
topology of such landscapes is important for understanding the
origins of the increasing number of laboratory successes in the
optimal control of quantum processes. This paper proposes a simple
scheme to compute the characteristics of the critical topology of
the quantum ensemble control landscapes, showing that the set of
disjoint critical submanifolds one-to-one corresponds to a finite
number of contingency tables that solely depend on the degeneracy
structure of the eigenvalues of the initial system density matrix
and the observable whose expectation value is to be maximized. The
landscape characteristics can be calculated as functions of the
table entries, including the dimensions and the numbers of
positive and negative eigenvalues of the Hessian quadratic form of
each of the connected components of the critical submanifolds.
Typical examples are given to illustrate the effectiveness of this
method.
\end{abstract}

\maketitle

\section{Introduction}
The control of physical and chemical quantum mechanical processes
has \cite{brixner,vlasta} recently seen many laboratory successes,
especially utilizing ultrafast shaped laser pulses as controls. In
many experiments, adaptive learning algorithms are applied to seek
optimal control fields \cite{JudRab1992,rabitz2,herek}. These
achievements collectively reveal a surprising ease in discovering
effective control fields despite the presence of noise,
imperfections and of severe constraints on the controls (i.e.,
limited bandwidth of $\sim$20nm operating at $\sim$800nm central
wavelength in virtually all of the experiments). This behavior
suggests that some generic foundation must lie behind the ease of
teaching a laser to guide quantum system dynamics to produce
specified outcomes. Recently, the notion of quantum control
landscapes was put forth to address this matter, where the
landscape is the map from the space of control fields to some
physical objective (e.g., quantum state transition probability
\cite{rabitz6,Baum2005}, the expectation value of a quantum
observable \cite{mike,tak-san}, or the fidelity of quantum gates
\cite{RabMik2005}). The landscape critical topology (i.e., the
topology of the set of critical points of the landscape) was
analyzed to give insight into the effort to search for an optimal
control. The initial exploration \cite{rabitz6} of the landscape
for the control of the state-to-state transition probability found
that the landscape had only perfect extrema such that no false
traps (i.e., local sub-optima in addition to the global optimum)
exist for the control searches to be caught in, which provides a
basic understanding of the ease of searching for optimal controls.

The same feature exists for generalized landscapes of the
observable expectation value on quantum ensembles
\cite{mike,tak-san} except for up to $\sim N!$ saddle suboptima,
where $N$ is the dimension of the system. The details of these
critical points depend on the observable and the system initial
state. For real quantum systems, especially molecular systems
whose electronic, vibrational and rotational states are involved,
it is essential to investigate the affects of degeneracies upon
the landscape topology to understand which classes of quantum
control problems are expected to be successfully treated in the
laboratory. Moreover, the landscape complexity needs to be
considered when analyzing the scaling of control field search
effort with $N$. This paper delves further into the control
landscape topology for general finite-level quantum systems with
arbitrary degeneracies in the statistical distribution of the
initial state and the spectral structure of the observable.

The paper is organized as follows. Section 2 presents a group
theoretical analysis of the quantum ensemble landscape in terms of
double cosets, as a generalization of torii obtained in
\cite{mike}. Section 3 ascribes the determination of double cosets
to the enumeration of contingency tables whose marginal
constraints are the degeneracy indices of the observable and the
system initial density matrix. Section 4 uses these contingency
tables to calculate several important topological features of the
landscape critical submanifolds and discusses the physical
implications of the findings. Section 5 provides some simple
illustrative examples, and conclusions are presented in Section 6.

\section{The Quantum ensemble control landscape}
For a $N$-level quantum system with non-dissipative dynamics, the
evolution of its density matrix is described by $\rho(t)=U(t)\rho
U^\dagger(t)$, where $\rho$ is the initial state and the unitary
system propagator $U(t)$ obeys the \sd equation
\begin{equation}\label{sd}
i\hbar \frac{\dd }{\dd t}U(t)=H[\ep(t)]U(t),~~~U(t_0)=I,
\end{equation}
Here the evolution extends over the interval $0\leq t\leq T$. The
governing Hamiltonian $H$ contains a time dependent control field
$\ep(t)$ that guides the evolution of the quantum system. The goal
is to find control fields that maximize the expectation value of
some desired quantum observable $\theta$ of the system at the time
$T$:
\begin{equation}\label{je}
J[\ep(\cdot)]=\tr\Big\{U(T)\rho U^\dagger(T)\theta\Big\}.
\end{equation}
The optimization also can be defined for a non-Hermitian operator
$\theta$ \cite{glaser}, but here only the Hermitian cases will be
studied. The quantum control landscape is defined as the mapping
$J$: $ \ep(\cdot)\rightarrow \mathbb{R}$ from the set of all
admissible controls into the real values of $J$. Accordingly, the
optimal controls must be critical points of $J$ in
$\ep(\cdot)$-space. A main objective of the landscape analysis is to
identify the geometry of all possible critical points, as they
specify the accessible stationary values of $J$ upon variation of
the control. A special issue of concern is the determination of
whether false suboptimal critical point traps exist, as their
presence could limit attempts to identify where to reach the
absolute maximum of the landscape.

As the landscape over the space of control fields is complex to
analyze, a natural approach is to re-express the analysis on the
unitary group:
\begin{equation}\label{el}
J(U)=\tr(U\rho U^\dagger\theta),\quad U\in \U(N),
\end{equation}whose critical topology is easier to determine.
The relationship between (\ref{je}) and (\ref{el}) is manifested
by the chain rule:
$$\delta J=\langle\nabla J(U(T)),\delta U(T)\rangle,$$
where $\nabla J(U(T))$ is the gradient of $J$ at $U(T)$ in
$\U(N)$, and $\delta U(T)$ is the variation of $U(T)$ caused by a
control variation $\delta \ep(\cdot)$ over $[0,T]$. The bilinear
operation $\langle A,B\rangle=Re\tr(A^\dagger B)$ is a Riemannian
metric defined on $\U(N)$. A control $\ep(\cdot)$ is called
regular if any variation $\delta U(T)$ of the resulting $U(T)$ is
attainable by some admissible control variation, otherwise it is
called singular \cite{BonChy2003}. For any regular control, the
sufficient and necessary condition for $\delta J\equiv 0$ upon any
control variation is $\nabla J(U(T))=0$, which implies that it is
critical if and only if the corresponding $U(T)$ is critical for
(\ref{el}). Moreover, the optimality status (i.e., as a minimum,
maximum or a saddle point) of the regular critical points and
their corresponding $U(T)$ are identical (see proof in
\cite{rebing}). Singular controls can also become critical points
of (\ref{je}) when the corresponding gradient vector $\nabla
J(U(T))$ is orthogonal to all attainable variations $\delta U(T)$,
but in such cases $\nabla J(U(T))$ is not necessarily vanishing.
In this regard, the landscape critical points can be classified as
regular and singular critical points, respectively. For landscape
studies, it is more important to assess the universal properties
that are Hamiltonian-independent, which are essential for
understanding the mounting successes in the control of various
quantum systems. Hence we are mainly concerned with the set of
regular controls, for which the reduction from $\ep(\cdot)$ to
$U(T)$ preserves the critical topology from the $\ep(\cdot)$-space
to $\U(N)$. The influence of singular controls will be contained
in another work.

The necessary and sufficient condition for $U$ to be a critical
point of (\ref{el}) has been derived in \cite{mike,helmke}. The
basic concept is that, under the parametrization
$U(s,A)=e^{isA}U$, where $s\in\R$ and $A^\dagger=A$, of a
neighborhood of $U$ in $\U(N)$, $U$ is critical if and only if:
$$\frac{\dd}{\dd s}J[U(s,A)]\Big|_{s=0}=0,~~~~\forall~A^\dagger=A,$$
which gives \begin{equation}\label{forallA} \tr(iA[\theta,U\rho
U^\dagger])=0,~~~~\forall~A^\dagger=A,\end{equation}and this is
equivalent to
\begin{equation}\label{vec} [\theta,U\rho U^\dagger]=0,
\end{equation}
which can be used to describe the critical topology
\cite{rabitz6,mike,glaser}. For systems with non-degenerate $\rho$
and $\theta$, the critical submanifolds consist of $N!$ number of
$N$-torii embedded in the unitary group $\U(N)$ \cite{rabitz6}.
The presence of degeneracies in $\rho$ or $\theta$ will merge
these torii into a smaller number of larger critical submanifolds,
and their explicit description will be resolved in the following
sections.

\section{Characterization of the Critical Submanifolds}
Let $R$ and $S$ be the unitary transformations that diagonalize
$\rho$ and $\theta$ respectively, i.e.,
\begin{eqnarray*}
&&  \tilde\rho =R^\dagger\rho R=diag\{\lambda_1,\cdots,\lambda_1;\cdots;\lambda_r,\cdots,\lambda_r\},  \\
 && \tilde\theta =S^\dagger\theta
 S=diag\{\epsilon_1,\cdots,\epsilon_1;\cdots;\epsilon_s,\cdots,\epsilon_s\},
\end{eqnarray*}
where $\lambda_1>\cdots>\lambda_r$ are distinct eigenvalues of
$\rho$ with $n_1,\cdots,n_r$ multiplicities and
$\epsilon_1>\cdots>\epsilon_s$ are distinct eigenvalues of
$\theta$ with $m_1,\cdots,m_s$ multiplicities. The landscape
functional can be written as \begin{equation}\label{tr-J} J(\tilde
U)=\tr(U R\tilde \rho R^\dagger U^\dagger S \tilde \theta
S^\dagger)=\tr[(S^\dagger U R) \tilde\rho (S^\dagger U R)^\dagger
\tilde \theta ]=\tr(\tilde U \tilde\rho \tilde U^\dagger \tilde
\theta )
\end{equation} where the automorphism $\tilde U=S^\dagger U R$ also runs
over $\mathcal{U}(N)$. Thus, without loss of generality, we can
always assume that both $\rho$ and $\theta$ are diagonal.

\begin{theorem}
Let $\U(\n)$ be the product group $\U(n_1)\times\cdots\times
\U(n_r)$ where $\U(n_i)$ is the $n_i$-dimensional unitary group
acting on the eigenspace of $\lambda_i$, and
$\U(\mm)=\U(m_1)\times\cdots\times \U(m_s)$ is defined in the same
manner relevant to $\theta$. A unitary matrix $U$ is a critical
point of (\ref{tr-J}) if and only if it is in the double coset
$$\U(\n)\pi\U(\mm)=\left\{P\pi Q:~~P\in
\U(\mm),~Q\in\U(\n)\right\}$$ of some permutation matrix $\pi$.
\end{theorem}
{\bf Proof:} Suppose that $U$ is a critical point. According to
(\ref{vec}), $U$ must transform $\rho$ into block-diagonal form,
whose block lengths correspond to the degenerate subspace
dimensions of $\theta$. This block-diagonal matrix $U\rho
U^\dagger $ can be subsequently diagonalized by a $m_1\times
\cdots\times m_s$ block-diagonal unitary matrix $P\in \U(\mm)$.
Since unitary transformations will not alter the spectrum of
$\rho$, the resulting diagonal matrix can be always written as
$\pi^\dagger \rho \pi$ where $\pi$ is a permutation matrix that
reorders the diagonal elements of $\rho$. These operations can be
expressed as $P^\dagger U \rho U^\dagger P= \pi \rho \pi^\dagger$,
which is equivalent to
$$ \rho  = (\pi^\dagger P^\dagger U  ) \rho (\pi^\dagger P^\dagger U)^\dagger.$$
By definition, the matrix $Q= \pi^\dagger P^\dagger U$ must be an
element of the stabilizer of $\rho$ in $\mathcal{U}(N)$, which
from group theory is identified with $\U(\n)$. Thus, we obtain the
decomposition $U=P\pi Q$. Conversely, one can easily verify that
an arbitrary permutation matrix is critical, $J(P\pi Q)=J(\pi)$
and the critical condition (\ref{vec}) is satisfied for both $\pi$
and $P\pi Q$, implying that $P\pi Q$ is also critical producing
the same landscape value as $\pi$. End of proof.

Theorem 1 shows that the critical manifold is a union of double
cosets:
\begin{equation}\label{p}
\M=\bigcup_{\pi\in \p(N)}{\U(\mm)\pi \U(\n)},
\end{equation}
where $\p(N)$ denotes the permutation group over $N$ indices. The
set $\M_\pi=\U(\mm)\pi\U(\n)$ is the double coset of $\pi$ in
$\mathcal{U}(N)$ with respect to $\U(\mm)$ and $\U(\n)$. In the
special case that both $\rho$ and $\theta$ are non-degenerate
\cite{mike}, the stabilizer of $\rho$ and $\theta$ are both
products of $N$ one-dimensional unitary groups, i.e.,
$$\U(\mm)=\U(\n)=[\U(1)]^{N},$$
the critical manifold as the double cosets can be verified to
consist of $N!$ disjoint $N$-torii, each of which is labelled by a
permutation matrix. Generally, the occurrence of degeneracies in
$\rho$ and $\theta$ will ``merge" two torii $\M_\pi$ and
$\M_{\pi'}$ together if $\pi'$ happens to be in $\M_\pi$, thereby
reducing the number of original disjoint critical submanifolds,
but increasing their dimensions. These critical submanifolds can
be taken as the generalized ``torii". Let $\tilde P$ be the set of
all inequivalent partitions of $\p(N)$ with respect to the
equivalence relation ``$\sim$" on $\p(N)$, which is defined as
$\pi'\sim \pi$ if $\pi'\in\M_\pi$. The characterization of the
disjoint critical submanifolds can be identified as seeking all
inequivalent double cosets of the permutation matrices in
$\mathcal{U}(N)$ with respect to $\U(\n)$ and $\U(\mm)$
\begin{equation}\label{dc}
\M=\bigcup_{\tilde\pi\in{\tilde P}}{\U(\mm)\tilde\pi \U(\n)}.
\end{equation}

We now introduce the concept of a contingency table to simplify
the abstract description of a critical submanifold expressed as
the double coset of some permutation $\pi$. Suppose that both
$\rho$ and $\theta$ have their diagonal elements arranged in
decreasing order. Let $k_{ij}$ be the number of positions on the
diagonal where the eigenvalues $\lmd_i$ and $\epsilon_j$ appear
simultaneously after imposing the permutation $\pi$ on $\theta$.
The contingency table consists of these nonnegative {\it overlap
numbers} as arranged in Table \ref{tab1}, whose row sums are
$n_i$'s and column sums are $m_j$'s.

\begin{table}[h]
\caption{\label{tab1}Contingency table $\K$.}
\begin{indented}
\lineup \item[]\begin{tabular}{c|cccc} \br
  & $~m_1~$ & $~m_2~$ & $\cdots$ & $~m_s~$\cr \mr
$~n_1~$ & $k_{11}$ & $k_{12}$ & $\cdots$ & $k_{1s}$ \cr
 $n_2$ & $k_{21}$ & $k_{22}$ & $\cdots$ & $k_{1s}$ \cr
 $\vdots$ & $\vdots$ & $\vdots$ & $\ddots$ & $\vdots$ \cr
 $n_r$ & $k_{r1}$ & $k_{r2}$ & $\cdots$ & $k_{rs}$\cr \br
\end{tabular}
\end{indented}
\end{table}

For example, for a three-level quantum system where
$\rho=diag\{0.4,0.3,0.3\}$ and $\theta=diag\{0.4,0.4,0.2\}$, the
corresponding marginal constraints are $n_1=1$ and $n_2=2$ for row
sums; $m_1=2$ and $m_2=1$ for column sums. Suppose that a
permutation transformation $\pi$ exchanges the second and third
eigenvalues of $\theta$, i.e., $\theta'=\pi\theta
\pi^\dagger=diag\{0.4,0.2,0.4\}$. Then the overlap number of the
first eigenvalue $0.4$ of $\rho$ with the first eigenvalue $0.4$
of $\theta$ is $k_{11}=1$, which appears at the first position on
the diagonal of $\rho$ and $\theta'$; $k_{22}=1$ is the overlap
number of the second eigenvalue $0.3$ of $\rho$ with the second
eigenvalue $0.2$ of $\theta$, which can be seen from the second
diagonal elements of $\rho$ and $\theta'$. Similarly, $k_{21}=1$
and $k_{12}=0$. The resulting contingency table is shown in Table
\ref{tab2}.

\begin{table}[h]
\caption{\label{tab2}The contingency table for \\ the illustrative
three-level system.}
\begin{indented}
\lineup \item[]\begin{tabular}{c|cccc} \br
  & $~2~$ & $~1~$ \cr\mr
$~1~$ & $1$ & $0$ \cr
 $~2~$ & $1$ & $1$ \cr\br
\end{tabular}
\end{indented}
\end{table}

Every permutation matrix leads to a contingency table, but the
same contingency table may be produced from different permutation
matrices which are mutually equivalent with respect to the
relation ``$\sim$". Therefore, the contingency tables provide an
equivalent description of the critical submanifolds:

\begin{theorem}
Every critical submanifold of (\ref{el}) is uniquely determined by
a contingency table satisfying the above marginal conditions, and
vice versa. The critical submanifold corresponding to a
contingency table $\K$ can be expressed as the quotient set
\begin{equation}\label{cm}
\M_\K=\frac{\U(\mm)\times \U(\n)}{\U(\K)}, \quad {\rm where}\,\,\,
\U(\K)=\prod_{k_{ij}\neq 0}\U(k_{ij}).
\end{equation}
\end{theorem}
{\bf Proof:} Define $F_\pi(P,Q)=P\pi Q$, where $(P,Q)\in
\U(\mm)\times \U(\n)$, as the bilateral operation on $\pi$. The
set of critical points that are equivalent with $\pi$ is
characterized by the stabilizer $stab(\pi)$ of $F_{\pi}$ in
$\U(\mm)\times \U(\n)$, i.e., the set of matrix pairs $(U,V)\in
\U(\mm)\times \U(\n)$ such that $F_\pi(U,V)=U\pi V=\pi$. Hence,
the critical submanifold $\M_\pi$ can be identified as the
quotient set of $\U(\mm)\times \U(\n)$ divided by $stab(\pi)$,
which is isomorphic to the set $\U(\n)\cap \pi^\dagger \U(\mm)
\pi$ as follows
$$stab(\pi)=\{(\pi
V^\dagger\pi ^\dagger,V):\,\,\,V\in \U(\n)\cap \pi^\dagger \U(\mm)
\pi\}.$$

The intersection set $\U(\n)\cap \pi^\dagger \U(\mm) \pi$ is a Lie
subgroup of $\U(N)$ and can be decomposed into the product of
smaller unitary groups as
$\U(\K)=\U(k_{11})\times\cdots\times\U(k_{rs})$, where the
$k_{ij}$'s are the entries of the contingency table for $\pi$.
This leads to the expression (\ref{cm}). Therefore, every critical
submanifold can be uniquely determined by a contingency table. On
the other hand, given the contingency table shown in Table
\ref{tab1}, one can shuffle the originally ordered diagonal
elements of $\theta$ such that $k_{ij}$ of the $m_j$ eigenvalues
$\epsilon_j$ are moved to any of the $n_i$ positions where the
eigenvalues $\lmd_i$ of $\rho$ are located. Any of such shuffles
can be represented by a permutation operation corresponding to
this table. Hence, any contingency table corresponds to at least
one critical submanifold. In this manner, we prove the one-to-one
correspondence between critical submanifolds and contingency
tables. End of proof.

The analysis above provides an easy means to analyze the critical
topology of general quantum ensemble landscapes, by which the
evaluation of the critical submanifolds on continuous Lie groups
is reduced to a simple finite combinatorial problem solvable by a
computer. From now on, we use the contingency tables to label the
different branches of the critical submanifolds. The landscape
value of the critical submanifold $\M_\K$ corresponding to
contingency table $\K$ is
$$J(\K)=\sum_{i=1}^N\rho_{ii}\theta_{\pi(i)\pi(i)}=\sum_{i=1}^r\sum_{j=1}^sk_{ij}\lambda_i\epsilon_j,$$
where $\pi$ is some permutation matrix whose contingency table is
$\K$.

\section{Characteristics of the critical submanifolds}
The contingency tables are not only convenient for labelling the
critical submanifolds, but also powerful in identifying their
intrinsic topological characteristics that are important for the
performance of search algorithms seeking effective quantum optimal
control field. These characteristics include (1) the dimension,
$D_0(\K)$, for a given critical submanifold $\M_\K$, which
qualitatively reflects the size of the associated critical
submanifold (especially the maximum of the landscape) that is
crucial for assessing robustness to control field noise; (2) the
numbers of the positive and negative ``principal axis directions"
near a critical submanifold, $D_+(\K)$ and $D_-(\K)$
\cite{rabitz6,mike}, which can affect the path taken by the search
algorithm through the influence of the number of positive and
negative directions in the vicinity of sub-optimal regions.

The dimension $D_0(\K)$ in terms of the associated contingency
table $\K$ can be calculated in an easier way by (\ref{cm}), with
the fact that the dimension of $U(\ell)$ is $\ell^2$:
\begin{eqnarray}
  D_0(\K) &=& {\rm dim}[\U(\n)]+{\rm dim}[\U(\mm)]-{\rm
dim}[\U(\K)] \nonumber \\
   &=& \sum_{i=1}^r n_i^2+\sum_{j=1}^s m_j^2-\sum_{i=1}^r\sum_{j=1}^sk_{ij}^2. \label{d0}
\end{eqnarray}

Computationally, the characteristics $D_0(\K)$ (resp., $D_+(\K)$
and $D_-(\K)$) can be identified as the numbers of zero (resp.,
positive and negative) eigenvalues of the Hessian quadratic form
(HQF) at any $U\in\M_\K$, which determines the geometry in the
vicinity of $U$. The HQF is defined as the second-order term of
$A$ in the Taylor expansion of $J(e^{iA}U)$ at $U$, and it can be
easily obtained as follows:
$$\mathcal{H}(A)=\tr(AU \rho  U^\dagger A\theta-A^2U \rho  U^\dagger \theta).$$
Let $x_{\beta\gamma}$ and $y_{\beta\gamma}$ be the real and
imaginary parts of the $\beta\gamma$-th matrix elements of $A$,
which represent the coordinate variables in the tangent space of
$U$. Then the Hessian form can be further expanded as the
following sum \cite{mike}:
\begin{equation}\label{hqf}
\mathcal{H}(A)=-\sum_{1\leq\beta<\gamma\leq
N}(\lambda_\beta-\lambda_\gamma)(\epsilon_\beta-\epsilon_\gamma)(x^2_{\beta\gamma}+y^2_{\beta\gamma}),
\end{equation}
where the real numbers $\lambda_\beta$ and $\lambda_\gamma$
($\beta,\gamma=1,\cdots,N$) are the rearranged $N$ diagonal
elements of $\rho$ and $\theta$ after some permutation operation
whose contingency table is $\K$. Since there are always $N^2$
terms in (\ref{hqf}), the sum of the three indices should satisfy
$$D_0(\K)+D_+(\K)+D_-(\K)=N^2.$$

According to (\ref{hqf}), the index $D_+$ is twice the number of
$(\beta,\gamma)$ pairs for which
$(\lambda_\beta-\lambda_\gamma)(\epsilon_\beta-\epsilon_\gamma)<0$,
while $D_-$ is twice the number of $(\beta,\gamma)$ pairs for which
$(\lambda_\beta-\lambda_\gamma)(\epsilon_\beta-\epsilon_\gamma)>0$.
A critical submanifold is locally maximal (or minimal) if the
corresponding index $D_+=0$ (or $D_-=0$), i.e., the Hessian
eigenvalues are all negative (or positive). This happens only if the
magnitudes of the $\lmd_\beta$'s and the $\epsilon_\gamma$'s are in
the same (or opposite) order, which corresponds to a unique maximal
(or minimal) critical submanifold. Therefore, there exist no local
suboptima over the control landscape. Besides the absolute maximal
and minimal critical submanifolds, all other critical points are
saddles. In practice, the saddles will never form false traps for a
search algorithm to approach global optimal controls, although the
search may be slowed down when the algorithm runs in their
neighborhoods. This behavior is a strong support for the observed
relative ease of searching for optimal controls in quantum systems.

Now suppose that $\lambda_\beta=\lambda_i$ and
$\lambda_\gamma=\lambda_p$ ($1\leq i,p\leq r$);
$\epsilon_\beta=\epsilon_j$ and $\epsilon_\gamma=\epsilon_q$
($1\leq j,q\leq s$). By the nature of the contingency table, there
are $k_{ij}k_{pq}$ possibilities for this to happen. Since
$\lmd_1,\cdots,\lmd_r$ and $\epsilon_1,\cdots,\epsilon_s$ are both
in decreasing order, we have the relationship:
$$(\lambda_\beta-\lambda_\gamma)(\epsilon_\beta-\epsilon_\gamma)\gtrless 0\quad \Longleftrightarrow \quad (i-p)(j-q)\gtrless
0.$$ Hence, the $D_\pm(\K)$ indices are equal to the following
quadratic function of $\K$:
\begin{equation}\label{dpm}
D_\pm(\K)=2\sum_{(i-p)(j-q)\lessgtr0}k_{ij}k_{pq},
\end{equation}which, by taking $\K$ as a $r \times s$ integer matrix, can be further written in a compact form:
\begin{eqnarray}
  D_+(\K) &=& 2\tr({\bf J}_r\K {\bf J}_s\K^\dagger), \label{d+} \\
  D_-(\K) &=& 2\tr({\bf J}_r\K {\bf J}_s^\dagger\K^\dagger).
  \label {d-}
\end{eqnarray}
Here the entries of the index matrix ${\bf J}_t=\{\sigma_{ij}\}$
($1\leq i,j\leq t$) are in upper triangular form
$$\sigma_{ij}=\left\{
\begin{array}{cc}
  1, & i<j\,; \\
  0, & i\geq j\,.\\
\end{array}%
\right.$$

We have not found an explicit formula for counting the number of
critical submanifolds, which mainly counts the saddle critical
submanifolds because there is always only one maximal and minimal
critical submanifold. Generally, the breaking of the degeneracies
in $\rho$ and $\theta$ (e.g., when $\rho$ turns from a pure state
to a mixed state) increases the number of saddle critical
submanifolds, which may have adverse impacts on the search
algorithms by creating a tortured path to approach the global
optima. The examples in Section 5 analyze some special cases where
the number of critical submanifolds can be explicitly calculated.
For more complex situations, some good estimates have been found
in \cite{bender,gail}.

In real physical systems, there always exist disturbances or some
other factors that may destroy the degenerate structures of the
initial density matrix and observable operator. Since all the
topological landscape properties are uniquely determined by their
degeneracy degrees, even a very small perturbation may alter the
critical topology. Heuristically, supposing the eigenvalues of
$\rho$ and $\theta$ are well separated, the perturbation will
wrinkle the existing ``flat" critical submanifolds such that new
smaller ones emerge, increasing the number of critical
submanifolds up to $N!$, while retaining only two absolute
extrema. The submanifold dimensions and principal axis directions
are consequently different in these cases. Nevertheless, provided
that the perturbation is sufficiently small, e.g., compared with
the step size taken upon numerical or experimental searching for a
control, we expect that the structural changes of the critical
topology will have little influence on the optimization algorithm
performance, because the original critical regions will remain
almost flat. To obtain a full view of the landscape, the
geometrical curvature that reflects ``steepness" (or ``flatness")
of the landscape near the critical submanifolds should also be
considered.

\section{Examples}\label{example}
In this section we will apply the results obtained above to
several simple examples. to illustrate some basic features of the
control landscape topology.

\subsection{Pure state system} Suppose that $\rho$ is a pure state, then $n_1=1$ for $\lambda_1=1$ and
$n_2=N-1$ for $\lambda_2=0$; the observable $\theta$ has $r$
distinct eigenvalues $\epsilon_1>\cdots>\epsilon_s$ with
degeneracies being $m_1,\cdots,m_s$. The enumeration of the
contingency tables is rather simple because the positions in only
the first row can be filled with a single $1$ and $s-1$ zeros, which
amounts to $s$ different contingency tables as shown in Table 2.
Therefore, there are $s$ distinct critical submanifolds for such
pure-state systems. Applying (\ref{d0}) and (\ref{dpm}), the
dimension and $D_\pm(\K_j)$ indices of the $j$-th critical
submanifold are
\begin{eqnarray*}
  D_0(\K_j) &=& N(N-2)+2m_j, \\
  D_+(\K_j) &=& 2(m_1+\cdots+m_{j-1}), \\
  D_-(\K_j) &=& 2(m_{j+1}+\cdots+m_{s}),
\end{eqnarray*} for $j=1,\cdots,s$. One can specify that $\M_{\K_1}$ is the maximum submanifold with
$$D_0(\K_1)=N(N-2)+2m_1,\quad D_-(\K_1)=2(N-m_1);$$
$\M_{\K_s}$ is the minimum submanifold with
$$D_0(\K_s)=N(N-2)+2m_s, \quad D_+(\K_s)=2(N-m_s).$$
Obviously, high degeneracy in the largest eigenvalue of the
observable may facilitate the optimal searches towards maximizing
the cost functional and enhance the robustness of perfect control,
while that of the smaller eigenvalues retards the control
searching on level sets of lower landscape values.

\begin{table}[h]
\caption{\label{tab3}Contingency tables $\K_j$ for a pure \\ state
system, $j=1,\cdots,s$.}
\begin{indented}
\lineup \item[]\begin{tabular}{c|ccccc} \br
   & $~m_1~$ & $\cdots$ & $~m_j~$ & $\cdots$ & $~m_s~$ \cr \mr
  1 & 0 & $\cdots$ & 1 & $\cdots$ & $0$  \\
  $N-1$ & $m_1$ & $\cdots$ & $m_j-1$ & $\cdots$ & $m_s$\\
  \br
\end{tabular}
\end{indented}
\end{table}

\subsection{Non-degenerate observable} Suppose the spectrum of $\theta$ is fully non-degenerate, i.e.,
$m_1=\cdots=m_N=1$; the density matrix $\rho$ has $r$ distinct
eigenvalues $\lmd_1>\cdots>\lmd_r$ with degeneracies being
$n_1,\cdots,n_r$. Then, the entries in the contingency table can
only be $0$ or $1$, and there are $\frac{N!}{n_1!(N-n_1)!}$ choices
for placing $n_1$ numbers $1$ in the $N$ positions in the first row.
Having fixed the first row, there are
$\frac{(N-n_1)!}{n_2!(N-n_1-n_2)!}$ choices to place $n_2$ numbers
$1$ in the remaining $N-n_1$ positions in the second row, etc.
Finally, the total number of distinct critical submanifolds is
$$\mathcal{N}=\frac{N!}{n_1!(N-n_1)!}\cdot\frac{(N-n_1)!}{n_2!(N-n_1-n_2)!}\cdots\frac{(N-n_1-\cdots-n_{r-2})!}{n_{r-1}!n_r!}=\frac{N!}{n_1!\cdots n_r!}.$$
The corresponding dimension of each critical submanifold is
$$D_0(\K_\alpha)=\sum_{i=1}^rn_i^2,\quad
\alpha=1,\cdots,\N.$$ From (\ref{cm}), these critical submanifolds
are diffeomorphic to each other. This result generalizes the
special case of both $\rho$ and $\theta$ being non-degenerate,
whose critical submanifolds are all $N$-torii. The $D_\pm(\K)$
indices vary with the corresponding contingency tables, among
which $D_-=N^2-\sum_i n_i^2$ at the maximum and $D_+=N^2-\sum_i
n_i^2$ at the minimum.

\subsection{A multi-degenerate system}
Consider an eight-level system with degeneracies $\n=(1,3,4)$ and
$\mm=(2,6)$. All of the possible contingency tables are listed as
below:

\begin{equation*}
\K_1=  \left[%
\begin{array}{cc}
  0 & 1 \\
  0 & 3 \\
  2 & 2 \\
\end{array}%
\right], \K_2= \left[%
\begin{array}{cc}
  0 & 1 \\
  1 & 2 \\
  1 & 3 \\
\end{array}%
\right],  \K_3=  \left[%
\begin{array}{cc}
  1 & 0 \\
  0 & 3 \\
  1 & 3 \\
\end{array}%
\right], \end{equation*}\begin{equation*} \K_4= \left[
\begin{array}{cc}
  0 & 1 \\
  2 & 1 \\
  0 & 4 \\
\end{array}
\right], \K_5= \left[%
\begin{array}{cc}
  1 & 0 \\
  1 & 2 \\
  0 & 4 \\
\end{array}%
\right].
\end{equation*}
The topological characteristics of the five critical submanifolds
are summarized in Table \ref{tab4}. The second critical
submanifold is the largest with dimension 50. The minimum
submanifold is the second largest. The maximum submanifold is the
smallest, and from (\ref{cm}), it is diffeomorphic to the fourth
saddle submanifold because they have the same group of nonzero
entries in their contingency tables. The high degree of
degeneracies yields the indicated high dimensions of the critical
submanifolds.

\begin{table}[h]
\caption{\label{tab4}Characteristics for the case of an
eight-level system with degeneracies $\n=(1,3,4)$ and
$\mm=(2,6)$.}
\begin{indented}
\lineup \item[]\begin{tabular}{c|ccccc} \br
 No.   & 1 &2 &3 & 4 & 5 \\ \mr
 Manifold dimension  &  48 & 50 & 46 & 44 & 44  \\
  Positive axis direction & 16 & 8 &  6 &  4 & 0  \\
 Negative axis direction & 0 & 6 & 12 & 16 & 20  \\
 Type &  ~minimum~ & ~saddle~ & ~saddle~  & ~saddle~ & ~maximum~  \\
\br\end{tabular}
\end{indented}
\end{table}

\subsection{Molecular systems}
A typical molecular system at room temperature and under strong
field bound-state control usually possesses a very large number
$N$ of accessible discrete levels. Suppose the system has
initially populated several non-degenerate vibrational levels
$|v_1\rangle,\cdots,|v_r\rangle$, each of which has associated
rotational states. Often the rotational states are densely packed,
and for illustration here, they will be taken as degenerate with
$n_i$ ($i=1,\cdots,r$) associated with each vibrational state
$|v_i\rangle$. The population $p_i$ of $|v_i\rangle$ is then
equally distributed over the $n_i$ rotational states. As the
control field can import considerable energy into the molecule,
$N$ can be very large such that $\sum_in_i\ll N$.

Consider the transition control from the initial state to a
$M$-dimension subspace, where both $M$ and $N-M$ are larger than
$\sum_in_i$. The target observable is expressed as a projector
$$\theta=\frac{1}{M}\sum_{i=1}^M |q_i\rangle\langle q_i|,$$
on this $M$-dimensional subspace. The contingency tables
corresponding to the critical submanifolds have to satisfy $(r+1)$
column-sum conditions $n_1,\cdots,n_r,N_0$, where
$N_0=N-\sum_{i=1}^r{n_i}$, and two row-sum conditions $M$ and
$N-M$, as shown in Table \ref{tab5}.
\begin{table}[h]
\caption{\label{tab5}Contingency tables for the \\ molecular
illustration.}
\begin{indented}
\lineup \item[]\begin{tabular}{c|ccccc} \br
   & $~n_1~$ & $\cdots$ & $~n_r~$ & $~N_0~$ \\ \mr
  $M$ &  $ k_{11}$&$\cdots$ & $k_{1r}$ & $k_{1,r+1}$  \\
  $N-M$ & $k_{21}$ & $\cdots$ & $k_{2r}$ & $k_{2,r+1}$\\
\br\end{tabular}
\end{indented}
\end{table}

Since the contingency table is determined by the values of
independent variables $k_{11},\cdots,k_{1r}$ which can vary from
$0$ to $n_i$ without violating the marginal condition, we can
obtain the following number of critical submanifolds
$$\N=\prod_{i=1}^r(n_i+1).$$
The global maximal submanifold corresponds to $k_{1i}=n_i$,
$i=1,\cdots,r$, which from (\ref{d0}) has dimension
$$D_0 = N^2-2(N-M)(n_1+\cdots+n_r).$$

Another special case is control to a specific final degenerate
state. For simplicity, we assume that the degeneracies of the
initially populated vibrational states are identical, i.e.,
$n_1=\cdots=n_r=n$. The projector $\theta$ then acts on a $n$
dimensional subspace, i.e., $M=n$. In this case, the number of
critical submanifolds equals the number of nonnegative unordered
partitions of $n$ in the first row of Table 4, which gives
$$  \N = \frac{(n+r)!}{n!r!}.$$
In this case, the dimension of the global maximal manifold is
$D_0=(N-n)^2+n^2$.

The molecular illustration above has (i) a small number of
initially populated states, (ii) a target population with a modest
number of states and (iii) a very large number $\N$ of accessible
states. These circumstances can arise quite commonly in molecular
control, and they produce the outcome that the global maximum
submanifold dimension $D_0$ is very large and scales as $\sim
N^2$. This behavior implies that finding a control on the maximum
critical submanifold should be relatively easy due to the large
size of the target submanifold and the lack of false traps.

\section{Discussion}
We have given a complete description of the critical topology of
quantum ensemble landscapes for general finite-level systems over
their propagator spaces. It is shown that the most important
topological features can be calculated from a set of contingency
tables that uniquely specify the critical submanifolds. This
framework opens up the possibility to explore the landscape of
more complex quantum systems. We have applied it to the control
landscape of open quantum systems \cite{rebing}, and another
potential application in the future is the analyses of control
landscapes for an infinite dimensional quantum system as the limit
of a series of finite dimensional systems, which can be dealt with
using the techniques developed here.

The landscape topology is based on a strong assumption that the
system is fully controllable, which may be limited in practice. In
particular, a broad class of partially controllable systems have
their propagators constrained on a proper subgroup $\G$ of
$\U(N)$, and the condition for some $U\in\G$ to be critical is
reduced to
 \begin{equation}\label{forsomeA} \tr(iA[\theta,U\rho
U^\dagger])=0,~~~~\forall~A\in{\bf g},\end{equation}where $\bf g$
is the Lie algebra of $\G$. Obviously, every critical point of the
landscape on $\U(N)$ that belongs to $\G$ still remains critical
on $\G$ (there may exist additional critical points since the
condition (\ref{forsomeA}) is weaker than (\ref{forallA})). In
certain cases, these critical points may degenerate from the
original saddle points into local maximal (minimal) points. Hence,
the results obtained here for fully controllable systems provide a
basis for future landscapes studies of uncontrollable systems.

\ack The authors acknowledge support from the DOE.

\newpage
\bibliographystyle{unsrt}
%\bibliography{hotice}

\begin{thebibliography}{10}

\bibitem{brixner}
T.~Brixner, N.H. Damrauer, G.~Krampert, P.~Niklaus, and G.~Gerber.
\newblock Femtosecond learning control of quantum dynamics in gases and
  liquids: Technology and applications.
\newblock {\em J. Modern Opt.}, 50:539, 2003.

\bibitem{vlasta}
V.~Bonacic-Koutechy and R.~Mitric.
\newblock Theoretical exploration of ultrafast dynamics in atomic clusters:
  analysis and control.
\newblock {\em Chem. Rev.}, 105:11--65, 2005.

\bibitem{JudRab1992}
R.S Judson and H.~Rabitz.
\newblock Teaching lasers to control molecules.
\newblock {\em Phys. Rev. Lett.}, 68:1500, 1992.

\bibitem{rabitz2}
H.~Rabitz, R.~de~Vivie-Riedle, M.~Motzkus, and K.~Kompa.
\newblock Whither the future of controlling quantum pheonomena?
\newblock {\em Science}, 288:824--828, 2000.

\bibitem{herek}
J.L. Herek, W.~Wohlleben, R.J. Cogdell, D.Zeidler, and M.~Motzkus.
\newblock Quantum control of energy flow in light harvesting.
\newblock {\em Nature}, 417:533, 2004.

\bibitem{rabitz6}
H.~Rabitz, M.~Hsieh, and C.~Rosenthal.
\newblock Quantum optimally controlled transition landscapes.
\newblock {\em Science 303 (2004):}, 303:1998--2001, 2004.

\bibitem{Baum2005}
M.~Wollenhaupt, A.~Prakelt, C.~Sarpe-Tudoran, D.~Liese, and
T.~Baumert.
\newblock Quantum control and quantum control landscapes using intense shaped
  femtosecond pulses.
\newblock {\em J. Mod. Opt.}, 52:2187--2195, 2005.

\bibitem{mike}
M.~Hsieh, R.~Wu, and H.~Rabitz.
\newblock The topology of the optimal control landscape for quantum ensemble
  preparation.
\newblock {\em submitted to J. Chem. Phys.}, 2007.

\bibitem{tak-san}
Tak-San Ho and H.~Rabitz.
\newblock Why do effective quantum controls appear easy to find?
\newblock {\em J. Photochemistry Photobiology A}, 180:226--240, 2006.

\bibitem{RabMik2005}
H.~Rabitz, M.~Hsieh, and C.~Rosenthal.
\newblock The landscape for optimal control of quantum-mechanical unitary
  transformations.
\newblock {\em Physical Review A}, 72:52337, 2005.

\bibitem{glaser}
S.J. Glaser, T.~Schulte-Herbruggen, M.~Sieveking, O.~Schedletzky,
N.C. Nielsen,
  O.W. Sorensen, and C.~Griesinger.
\newblock Unitary control in quantum ensembles: maximizing signal intensity in
  coherent spectroscopy.
\newblock {\em Science}, 280:421--424, 1998.

\bibitem{BonChy2003}
B.~Bonnard and M.~Chyba.
\newblock {\em Singular trajectories and their role in control theory},
  volume~40 of {\em Mathematiques and Applications}.
\newblock Springer, Berlin, 2003.

\bibitem{rebing}
R.~Wu, A.~Pechen, H.~Rabitz, M.~Hsieh, and B.~Tsou.
\newblock Control landscapes for observable preparation with open quantum
  systems.
\newblock {\em submitted to J. Math. Phys.}, 2007.
\newblock Preprint available at http://arxiv.org/abs/0708.2119.

\bibitem{helmke}
U.~Helmke and J.~B. Moore.
\newblock {\em Optimization and dynamical systems}.
\newblock Springer-Verlag, London, 1994.

\bibitem{bender}
E.A. Bender.
\newblock The asymptotic number of non-negative integer matrices with given
  row and column sums.
\newblock {\em Discrete Math.}, 10:217--223, 1974.

\bibitem{gail}
M.~Gail and N.~Mantel.
\newblock Counting the number of $r\times c$ contingency tables with fixed
  margins.
\newblock {\em J. Amer. Statist. Assoc.}, 72:859--862, 1977.

\end{thebibliography}

\end{document}